\documentclass[letterpaper,twocolumn,10pt]{article}
\usepackage{usenix-2022}

\usepackage{amsmath}
\usepackage{xspace}
\usepackage{titling}
\usepackage[small,compact]{titlesec}

\usepackage{booktabs}
\usepackage{multirow}

\usepackage{graphicx}
\graphicspath{ {figures/} }
\usepackage{enumitem}
\usepackage{textcomp}
\usepackage{etoolbox}
\usepackage{xcolor}
\usepackage[T1]{fontenc}

\newtoggle{showmarks}
\toggletrue{showmarks}

\newcommand{\sn}{Apiary\xspace}
\newcommand{\algoname}{SFR\xspace}

\usepackage{caption}
\usepackage{subcaption}     
\usepackage{comment}
\usepackage{fancyvrb}

\usepackage{textcomp} 

\usepackage{algorithm}
\usepackage[noend]{algpseudocode}
\makeatletter
\renewcommand{\ALG@beginalgorithmic}{\small}
\makeatother
\usepackage{eqparbox}
\newdimen{\algindent}
\algnewcommand\LeftComment[2]{%
\hspace{#1\algindent}$\triangleright$ \eqparbox{COMMENT}{#2} \hfill %
}

\usepackage{listings}
\definecolor{codegreen}{rgb}{0,0.6,0}
\definecolor{codeblue}{rgb}{0,0.5,1.0}
\definecolor{codegray}{rgb}{0.4,0.4,0.4}
\definecolor{codepurple}{rgb}{0.58,0,0.82}
\definecolor{backcolour}{rgb}{0.95,0.95,0.92}

\lstdefinestyle{mystyle}{
    commentstyle=\color{codegray},
    keywordstyle=\color{magenta},
    numberstyle=\tiny\color{codegray},
    stringstyle=\color{codepurple},
    basicstyle=\ttfamily\small,
    breakatwhitespace=false,
    breaklines=true,
    captionpos=b,
    keepspaces=true,
    numbers=left,
    numbersep=5pt,
    showspaces=false,
    showstringspaces=false,
    showtabs=false,
    tabsize=2,
    xleftmargin=3mm,
    numberblanklines=false,
    escapeinside=||
}
\newcommand*\DNumber{\addtocounter{lstnumber}{-1}}
\usepackage{enumitem}
\usepackage{tcolorbox}
\setitemize{topsep=2pt, itemsep=0em, leftmargin=1em}
\setenumerate{topsep=2pt, itemsep=0em, leftmargin=1.4em}

\vfuzz \hfuzz

\begin{document}
\date{}

\setlength{\droptitle}{-0.55in}

\title{\Large \bf \sn: A DBMS-Integrated Transactional Function-as-a-Service Framework
\vspace{-0.25in}
}

\author{Peter Kraft$^{1*}$, Qian Li$^{1}$\thanks{Both authors contributed equally to this paper.}, Kostis Kaffes$^{1}$, Athinagoras Skiadopoulos$^1$, Deeptaanshu Kumar$^3$, 
Danny Cho$^1$, \\
Jason Li$^2$, Robert Redmond$^2$, Nathan Weckwerth$^2$, Brian Xia$^2$, Peter Bailis$^1$, Michael Cafarella$^2$, \\
Goetz Graefe$^4$, Jeremy Kepner$^2$, Christos Kozyrakis$^1$,
Michael Stonebraker$^2$, Lalith Suresh$^5$, \\
Xiangyao Yu$^6$, and Matei Zaharia$^1$
\\
  $^1$Stanford, $^2$MIT, $^3$CMU, $^4$Google, $^5$VMware, $^6$University of Wisconsin-Madison
}

\maketitle

\begin{abstract}
Developers increasingly use function-as-a-service (FaaS) platforms for \emph{data-centric} applications that perform low-latency and transactional operations on data, such as for microservices or web serving.
Unfortunately, existing FaaS platforms support these applications poorly because they physically and logically separate application logic, executed in cloud functions, from data management, done in interactive transactions accessing remote storage.
Physical separation harms performance while logical separation complicates efficiently providing transactional guarantees and fault tolerance.

This paper introduces \sn, a novel DBMS-integrated FaaS platform for deploying and composing fault-tolerant transactional functions.
\sn physically co-locates and logically integrates function execution and data management by wrapping a distributed DBMS engine and using it as a unified runtime for function execution, data management, and operational logging, thus providing similar or stronger transactional guarantees as comparable systems while greatly improving performance and observability.
To allow developers to write complex stateful programs, we leverage this integration to enable efficient and fault-tolerant function composition, building a frontend for orchestrating workflows of functions with the guarantees that each workflow runs to completion and each function in a workflow executes exactly once.
We evaluate \sn against research and production FaaS platforms and show it outperforms them by 2--68$\times$ on microservice workloads by reducing communication overhead.
\end{abstract}

\section{Introduction}
\label{sec:introduction}

Function-as-a-service (FaaS), or serverless, cloud offerings are becoming popular in both industry and research applications~\cite{jonas2019cloud}.
In widely-used FaaS platforms like AWS Step Functions~\cite{stepfunctions} and Azure Durable Functions~\cite{durablefunctions}, developers write programs as workflows of stateless functions whose deployments are managed by the service provider.
FaaS radically reduces the operational complexity of cloud deployment by eliminating the need to manage application servers.

\begin{figure}[t!]
	\centering
	\includegraphics[width=0.95\linewidth]{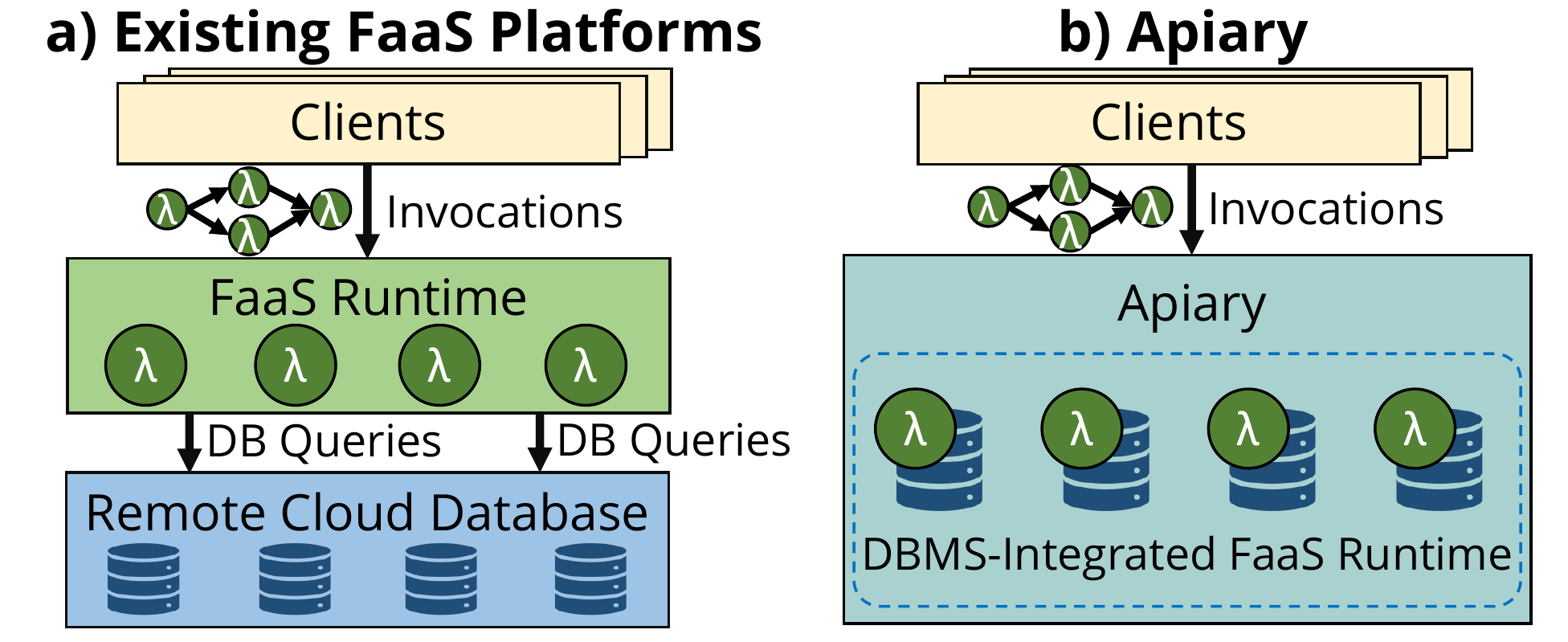}
	\caption{
		Existing FaaS platforms separate application logic,
		executed in cloud functions, 
		from data management, done in interactive transactions accessing a remote database.
		\sn instead tightly integrates application logic and data management,
		executing functions in DBMS stored procedures.
	}
	\label{fig:intro_figure}
\end{figure}

\begin{figure}[t!]
	\centering
	\includegraphics[width=0.9\linewidth]{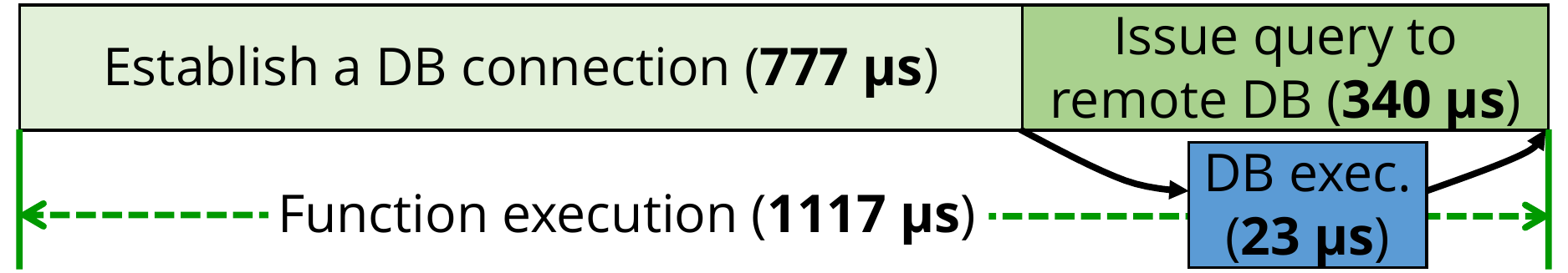}
	\caption{
		Latency breakdown for an OpenWhisk function performing a point database update.
		Query execution accounts for only 2\% of the overall function execution time.
	}
	\label{fig:ow_latency_breakdown}
\end{figure}

The FaaS model is increasingly popular for data-centric applications:  low-latency and transactional applications such as an e-commerce web service.
Unfortunately, existing FaaS platforms support these applications poorly because they both physically and logically separate function execution from data management, adopting the architecture of Figure~\ref{fig:intro_figure}a and calling a remote DBMS once per data operation.
Physical separation causes high communication overhead: in our experiments with OpenWhisk (Figure~\ref{fig:ow_latency_breakdown}), communication accounts for as much as 98\% of function runtime.
Logical separation complicates fault tolerance and transactional guarantees because functions may be arbitrarily re-executed and transactions cannot span across functions.
There have been several attempts to tackle these problems, such as~\cite{beldi,cloudburst,boki,de2021distributed,shredder}, providing either physical co-location for good performance or logical integration for strong gurantees, but not both.
Some, like Cloudburst~\cite{cloudburst}, physically co-locate compute and data using local caches,
but do not provide transactions.
Others, like Boki~\cite{boki} and Beldi~\cite{beldi}, provide transactional functions using an external transaction manager over remote storage, increasing already-high storage access times by 3$\times$.

In this paper, we present \sn, a transactional, high-performance FaaS platform for data-centric applications.
Unlike existing platforms, \sn physically co-locates and logically integrates function execution and data management by wrapping a distributed DBMS engine and using it as a unified runtime for function execution, data management, and operational logging (Figure~\ref{fig:intro_figure}b).
We compile functions to stored procedures, routines in a non-SQL language that run natively as DBMS transactions, thus making functions basic units of both control flow and atomicity.
We then leverage this integration to support efficient, fault-tolerant function composition and new observability capabilities.
We demonstrate that \sn provides \textbf{(a)} similar or stronger guarantees than comparable platforms, \textbf{(b)} performance improvements of 2--68$\times$ compared to state-of-the-art systems, and \textbf{(c)} automatic tracing of application-database interactions for observability.

The major challenge in designing a FaaS platform for data-centric applications is providing efficient, fault-tolerant function composition.
FaaS developers write complex programs by composing functions into \emph{workflows}.
To be robust to failures, workflows require strong execution guarantees: they must always run to completion and each of their functions must execute exactly once.
Existing platforms provide these guarantees by requiring functions to be idempotent~\cite{aws_idempotent} or building costly external transaction managers~\cite{beldi}.
By contrast, we can leverage \sn's integration of functions and data to build a fault-tolerant frontend providing these guarantees with minimal developer requirements and low overhead.
Because functions are stored procedures, \sn can instrument them to transactionally record their executions in the DBMS.
Therefore, if a workflow execution fails, the frontend can safely resume it by invoking its functions from the beginning, skipping functions that have already executed.
However, naively instrumenting all functions is expensive, degrading performance up to 2.2$\times$.
Thus, we develop a novel algorithm to identify when functions can be safely re-executed, reducing overhead to <5\%.

\sn also tackles a common challenge faced by FaaS developers:
obtaining observability into how applications interact with data.
Existing platforms can record function executions,
but it is hard for them to capture interactions with data
because they lack visibility into how functions manage data,
so developers must resort to expensive and error-prone manual logging across many functions.
\sn naturally has this visibility because it tightly integrates functions and data
and can leverage existing techniques for database provenance capture.
Therefore, we build a tracing layer that traces application control flow across functions through workflow instrumentation,
then records which data items each function accesses or updates through query instrumentation and change data capture
to produce a complete history of application interactions with data.
Obtaining this detailed information through manual logging incurs overhead of up to 92\%,
but our tracing layer reduces this to <15\% by building
a high-performance in-memory buffer and exporting its contents asynchronously. 

We evaluate \sn with commonly used microservice and web serving benchmarks
such as social networks and e-commerce sites~\cite{deathstarbench,hipster_shop}.
We show that by reducing communication overhead, it outperforms the popular open-source FaaS platform OpenWhisk~\cite{openwhisk} by 7--68$\times$ and recent research systems like Cloudburst~\cite{cloudburst} and Boki~\cite{boki} by 2--27$\times$.

In summary, our contributions are:

\begin{itemize}
	\item We propose \sn, a transactional FaaS platform that physically co-locates and logically integrates functions with data by wrapping a distributed DBMS.
	\sn outperforms research and production platforms by 2--68$\times$ while providing similar or stronger guarantees.
	
	\item We leverage \sn's architecture to design a fault-tolerant frontend for orchestrating workflows of functions.
	It guarantees that regardless of failures, workflows always run to completion and their functions each execute exactly once.
	
	\item We use \sn's architecture to enhance observability by automatically instrumenting applications and their interactions with data, achieving <15\% overhead compared to 92\% with manual logging.
\end{itemize}

\section{\sn Overview}
\label{sec:system_overview}

\subsection{System Architecture}
\label{sec:system-architecture}

\begin{figure}[t!]
	\includegraphics[width=\linewidth]{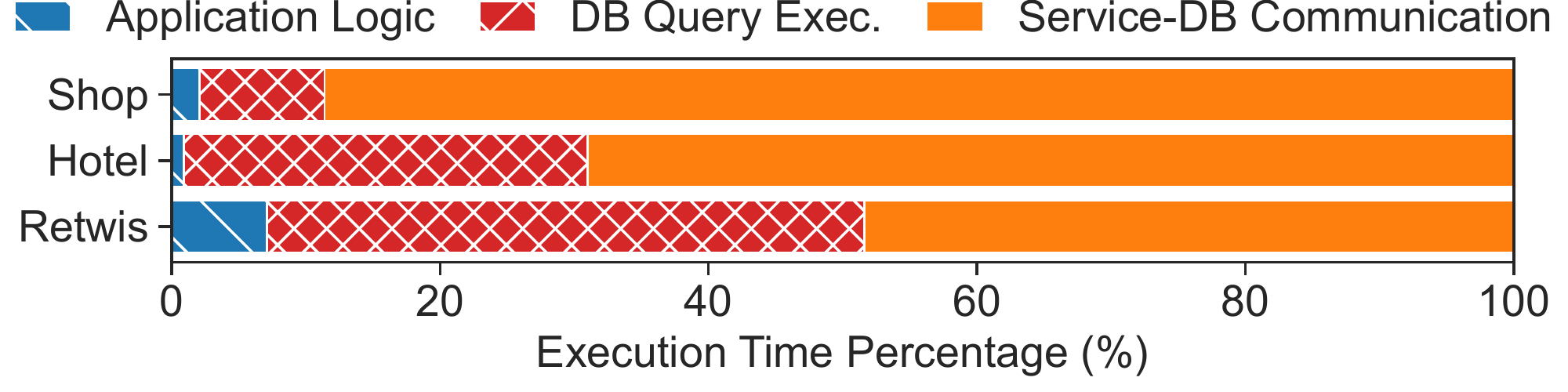}
	\caption{
		Average execution time breakdown for three data-centric applications used in evaluation (\S\ref{sec:evaluation}).  All three spend most of their
		runtime communicating with the remote database (over long-lived connections) or executing database operations.
	}
	\label{fig:exec_time_breakdown}
\end{figure}

\sn's design is motivated by a key observation: data-centric applications spend most of their runtime either communicating with a DBMS or executing DBMS operations.
As we show in Figure~\ref{fig:exec_time_breakdown}, these account for 93-99\% of the runtime of the microservice applications we use in our evaluation (\S\ref{sec:evaluation}).
To reduce communication overhead and improve performance, we architect \sn to physically co-locate compute and data by wrapping a distributed DBMS and compiling functions to database stored procedures.
Because stored procedures are transactional, this architecture also logically integrates function execution and data management; we leverage this integration to efficiently provide transactional guarantees (\S\ref{sec:semantics}), fault tolerance (\S\ref{sec:semantics}, \S\ref{sec:execution}), and observability (\S\ref{sec:provenance}).

\begin{figure}[t!]
	\centering
	\includegraphics[width=0.95\linewidth]{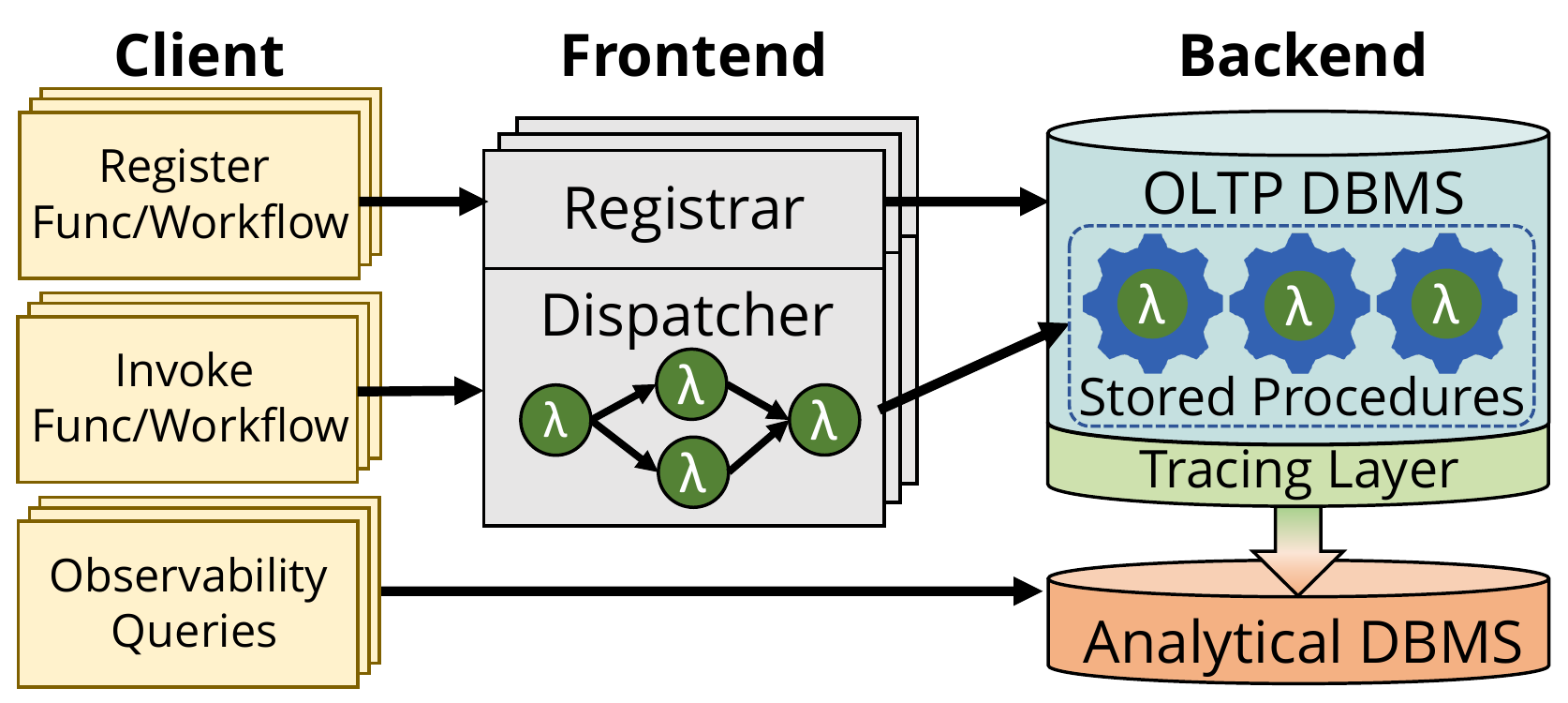}
	\caption{
		Architecture of \sn. 
	}
	\label{fig:system_architecture}
\end{figure}

\sn's architecture assumes developers write programs as workflows of functions; we discuss the programming interface and its semantics in \S\ref{sec:semantics}.
We sketch the architecture in Figure~\ref{fig:system_architecture}.
It has three layers: the clients, frontend, and backend.

\noindent\textbf{Clients.}
Clients send requests through a client library to the frontend to invoke workflows and functions, which execute in the backend.
Developers write functions and compose workflows using our programming interface.

\noindent\textbf{Frontend.}
Frontend servers route and authenticate client requests to the backend.
Each server has a dispatcher, which manages workflow execution by invoking each function in the backend, passing in its inputs, collecting its outputs to send to later functions, and enforcing fault-tolerance guarantees (\S\ref{sec:semantics}, \S\ref{sec:execution}).
Servers also contain registrars, which handle function and workflow registration, instrumentation, and compilation.

\noindent\textbf{Backend.}
The backend executes functions and manages data.
It wraps a distributed DBMS and its stored procedures.
\sn executes functions transactionally on DBMS servers, instrumenting them to provide fault tolerance (\S\ref{sec:semantics}, \S\ref{sec:execution}) and capture information on application-database interactions for observability (\S\ref{sec:provenance}).
Because functions are physically co-located with the DBMS, we rely on the DBMS's native elastic scaling capabilities to scale the backend.

\subsection{Non-Goals}

We want to emphasize two objectives that are \emph{excluded} from the scope of this paper.

\noindent\textbf{Compute-Heavy Workloads.}
\sn's design focuses on short-lived data-centric applications,
not long-running compute-intensive workloads such as video processing~\cite{excamera} or batch analytics~\cite{locus}.
These do not require its features and guarantees, such as transactional functions and exactly-once semantics.
If users wish to execute compute-heavy tasks,
we expect them to leverage an external service, such as AWS Rekognition~\cite{rekognition} for text detection in images.

\noindent\textbf{Non-Relational Data Models.}
\sn currently only supports a relational data model.
We believe it is possible to extend \sn to support transactional non-relational databases,
such as MongoDB, but this is beyond the scope of this paper.
Most comparable data-centric FaaS platforms are also restricted to relational or key-value data, e.g.~\cite{boki,cloudburst,faasm,shredder,beldi}.
\section{\sn Semantics}
\label{sec:semantics}

\sn aims to make it easy for developers to write stateful applications which are performant, robust to failures, and correctly handle concurrent operations.
To make this possible, we provide a familiar FaaS programming interface (\S\ref{sec:programming_model}) similar to widely-used platforms like AWS Step Functions~\cite{stepfunctions}, but offer much stronger guarantees, including transactional semantics (\S\ref{sec:transactional-semantics}) and fault tolerance (\S\ref{sec:fault-tolerance-semantics}).

\subsection{Programming Interface}
\label{sec:programming_model}

Before discussing \sn's semantics, we sketch its programming interface, using as an example a hotel reservation service that checks if a room is available, books it, then sends a confirmation email.
Developers write functions in a high-level language (Java) using SQL to access data stored in a relational DBMS, then construct programs as workflows of functions.

\begin{figure}[t]
	\includegraphics[width=\linewidth]{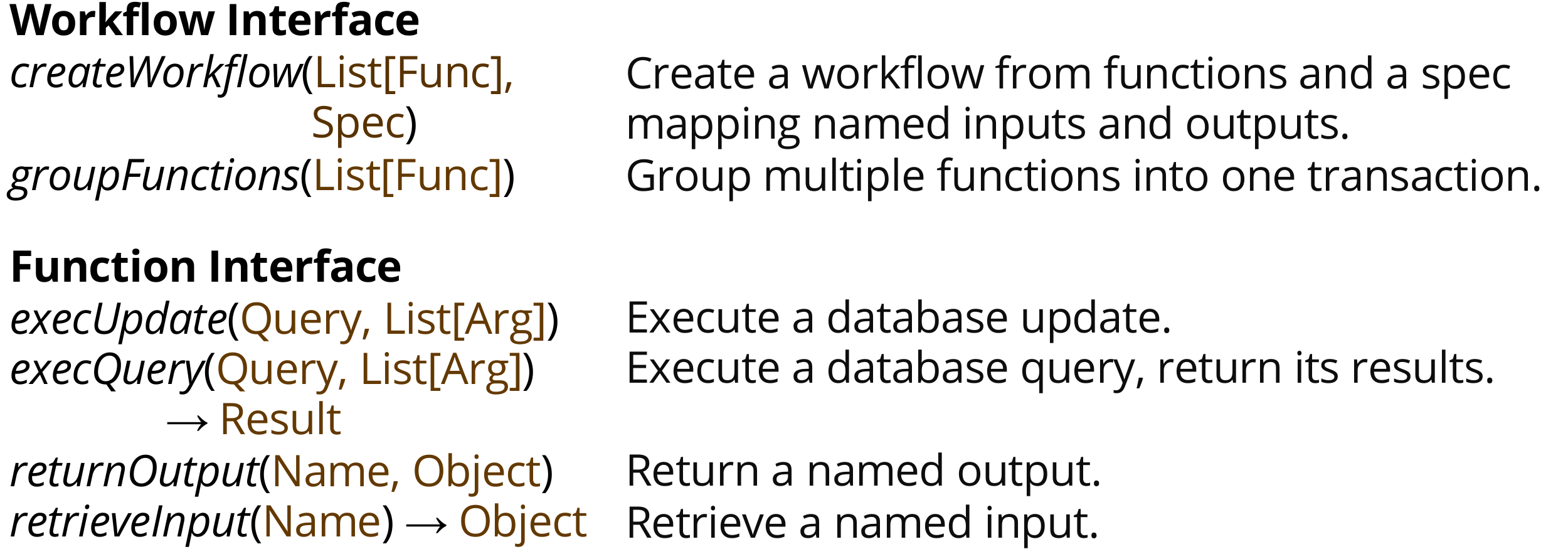}
	\caption{The \sn workflow and function interfaces.
	}
	\label{fig:dbos_interface}
\end{figure}

\paragraph{Function Interface.}
Functions take in and return any number of named serializable objects.
They can embed SQL queries to access or modify data in the DBMS.
We show the function interface in Figure~\ref{fig:dbos_interface}.
In the hotel reservation example, we implement checking availability, booking a room, and sending an email in separate functions; we show the code for checking availability in Figure~\ref{fig:example-code} (lines 1--10).

\sn requires functions to follow three rules:
\begin{enumerate}
	\item All SQL queries in functions must be defined statically as parameterized prepared statements.
	\item Functions must be deterministic.
	\item External service or API calls must be idempotent.
\end{enumerate}

\noindent The first rule enables static analysis (\S\ref{sec:execution}) and data tracing for observability (\S\ref{sec:provenance}); the last two rules enable practical implementation of exactly-once semantics (\S\ref{sec:execution}).

\paragraph{Workflow Interface.}
Developers construct programs as \emph{workflows} of functions using the interface in Figure~\ref{fig:dbos_interface}.
Each workflow is a directed acyclic graph (DAG) where nodes are functions, edges are data flow, and the input to a function is the output of its parents.
Developers construct a workflow from a list of functions and a specification mapping outputs of earlier functions to inputs of later functions.
Recursive or cyclic dependencies are not allowed.
Each workflow has a single \emph{sink} function that has no children; its output is returned to the client.
To guarantee the correctness of complex workflows, \sn lets developers designate a group of functions in a workflow as a single transaction; we discuss this in detail in \S\ref{sec:transactional-semantics}.
We can implement the hotel reservation service as a three-function workflow (\texttt{checkAvail}$\Rightarrow$\texttt{reserve}$\Rightarrow$\texttt{sendEmail}), where we group the first two functions into one transaction;
we sketch this in Figure~\ref{fig:example-code} (lines 12--15).

\begin{figure}[t]
\begin{lstlisting}[linewidth=\columnwidth,language=Java,basicstyle=\scriptsize\fontfamily{lmtt}\fontseries{m}\selectfont,tabsize=2,aboveskip=0pt,belowskip=0pt]
def checkAvail():
	query = new SQL("SELECT numAvail FROM HotelAvail WHERE hotelID=? AND date=?")
	inp = retrieveInput("availIn")
	avail = true;
	for (dt = inp.start; dt < inp.end; dt++):
		num = execQuery(query, inp.hotelID, dt)
		if (num < inp.numRooms):
			avail = false
			break
	returnOutput("availOut", avail)|\DNumber|

// Omit reserve and sendEmail due to space limit.
w = createWorkflow([checkAvail, reserve, sendEmail],
	{"in": "availIn", "availOut": "reserveIn",
	"reserveOut": "emailIn", "emailOut": "out"})
w.groupFunctions([checkAvail, reserve])
\end{lstlisting}
	\caption{Pseudocode implementing the three-function hotel workflow using the \sn interface (highlighted in \textcolor{blue}{blue}).}
	\label{fig:example-code}
\end{figure}

\subsection{Transactional Semantics}
\label{sec:transactional-semantics}

FaaS programs often require transactional guarantees; for example, our hotel reservation workflow needs transactions to guarantee that rooms are never double-booked.
Thus, \sn logically integrates function execution and data management: each function executes as a serializable ACID transaction in the database.
A key implication of this design is that functions are units of both control flow and atomicity; we leverage this to implement fault tolerance (\S\ref{sec:execution}) and track application-database interactions for observability (\S\ref{sec:provenance}).

An important question is what transactional semantics we provide for workflows.
Naively, we could execute entire workflows in a single transaction, but this leads to unnecessarily large transactions with poor performance.
Alternatively, we could provide no transactional guarantees for workflows and allow functions from concurrent workflows to arbitrarily interleave, but developers often require transactional guarantees across multiple functions.
For example, in the hotel workflow, the first two functions (\texttt{checkAvail} and \texttt{reserve}) must execute in one transaction to ensure the room is actually available when it is booked.
To balance these tradeoffs, we provide \emph{multi-function transactions}: developers can group multiple functions in a workflow to execute as a single ACID transaction, provided they form a connected subgraph of the workflow graph.
This design gives developers the flexibility to transactionally execute related operations, but separate unrelated operations to avoid the performance overhead of excessively large transactions.

\subsection{Fault-Tolerance Guarantees}
\label{sec:fault-tolerance-semantics}

Transactions are not sufficient to guarantee robust workflow execution in the presence of failures because they do not prevent issues such as workflows being executed partially or individual functions being executed multiple times.
Thus, \sn provides two guarantees for robust workflow execution.
First, workflows \emph{run to completion}, so even if the failure of a dispatcher or DBMS server causes workflow execution to halt, the workflow is eventually resumed.
Second, functions in workflows execute \emph{exactly once}, so even if a workflow or any of its functions fails and is restarted multiple times, the effect of the workflow on application state (in the database) is the same as if every function in the workflow were executed exactly once.
For example, in the hotel workflow, \sn guarantees that a room is only booked once and that if it is booked successfully, a confirmation email is always sent.

\subsection{Comparison with Related Systems}
\label{sec:semantics-comparison}

In Table~\ref{tab:semantics-comparison}, we compare the semantics of \sn to those of related systems.
\sn provides substantially stronger guarantees than commercial systems such as AWS Step Functions~\cite{stepfunctions} and Azure Durable Functions~\cite{durablefunctions}.
These support run-to-completion workflows but neither provide transactional function guarantees nor have visibility into the application database.
As a result, they cannot provide exactly-once function execution, instead providing the weaker at-least-once guarantee and allowing arbitrary function re-execution~\cite{aws_idempotent}.
AWS Step Functions claims it can also offer exactly-once semantics, but this is actually an at-most-once guarantee: it does not retry on failure, but instead guarantees tasks never run more than once~\cite{stepfunctionsemantics}.

\sn provides similar guarantees to transactional FaaS systems such as Beldi~\cite{beldi}, Boki~\cite{boki}, and Transactional Statefun~\cite{de2021distributed}.  
They all allow developers to provide ACID transactional guarantees for individual functions, though  Transactional Statefun implements a limited ``one-shot'' model where the outputs of earlier queries in a transaction cannot be used as inputs to later queries in the same transaction.
All provide exactly-once semantics for functions and run-to-completion workflows.
None provide transactional guarantees for entire workflows, but all support running multiple functions in a single transaction similar to \sn multi-function transactions.
In Beldi and Boki, a transactional function can synchronously call other functions so they all execute in one large transaction.
In Transactional Statefun, a ``coordinator function'' can coordinate multiple other functions through two-phase commit so they execute as a single transaction.
However, while these systems build costly external transaction managers over remote storage, \sn instead minimizes transactional overhead by co-locating compute and data.

An important related class of systems is FaaS platforms built around causal consistency, such as Cloudburst~\cite{cloudburst}, Hydrocache~\cite{hydrocache}, and FaaSTCC~\cite{faastcc}.  
These store data in a remote key-value store and use local caches to improve performance.
The strongest guarantee they provide is transactional causal consistency (TCC) for entire workflows.
TCC guarantees that workflows cannot see the effects of other workflows until they are entirely complete, a guarantee only provided by \sn or other transactional FaaS systems if the entire workflow is executed in a single multi-function transaction.
However, for individual functions, TCC provides relatively weak guarantees, allowing serious anomalies such as stale reads and write-write conflicts.
By contrast, \sn runs each function as an ACID transaction with serializable isolation, disallowing these anomalies.
Moreover, these systems only provide a key-value API for data management, while \sn supports a relational model.

\begin{table}[t!]
  \setlength{\tabcolsep}{3pt}
  \resizebox{\linewidth}{!}{%
  \centering
  \footnotesize
  \begin{tabular}{@{}llllll@{}}
      \toprule
      Platform  & \begin{tabular}[c]{@{}l@{}}Transactional\\ Functions\end{tabular} & \begin{tabular}[c]{@{}l@{}}Multi-Func.\\ Txns. \end{tabular} & \begin{tabular}[c]{@{}l@{}}Exactly-Once\\ Semantics\end{tabular} & \begin{tabular}[c]{@{}l@{}}Run-to-\\ Completion \end{tabular} & \begin{tabular}[c]{@{}l@{}}Data\\ Locality\end{tabular}\\
      \midrule
      Step Functions~\cite{stepfunctions}         & \textcolor{red!70!black}{No}   & \textcolor{red!70!black}{No}        & {\color[HTML]{583200}At-Least-Once} & \textcolor{green!50!black}{Yes}   & \textcolor{red!70!black}{No} \\
      Durable Functions~\cite{durablefunctions}  & \textcolor{red!70!black}{No}   & \textcolor{red!70!black}{No}        & {\color[HTML]{583200}At-Least-Once} & \textcolor{green!50!black}{Yes} & \textcolor{red!70!black}{No} \\ \midrule
      Cloudburst~\cite{cloudburst}         & {\color[HTML]{583200}CC}  & {\color[HTML]{583200}CC}       & \textcolor{red!70!black}{No}            & \textcolor{red!70!black}{No}  & {\color[HTML]{583200}Caching} \\
      FaaSTCC~\cite{faastcc}            & {\color[HTML]{583200}TCC} & {\color[HTML]{583200}TCC}       & {\color[HTML]{583200}At-Least-Once} & \textcolor{green!50!black}{Yes} & {\color[HTML]{583200}Caching} \\
      Hydrocache~\cite{hydrocache}         & {\color[HTML]{583200}TCC} & {\color[HTML]{583200}TCC}       & {\color[HTML]{583200}At-Least-Once} & \textcolor{green!50!black}{Yes}  & {\color[HTML]{583200}Caching} \\ \midrule
      StateFun-Txns~\cite{de2021distributed}      & \textcolor{green!50!black}{Yes}  & \textcolor{green!50!black}{Yes}       & \textcolor{green!50!black}{Yes}           & \textcolor{green!50!black}{Yes} & \textcolor{red!70!black}{No}  \\
      Beldi~\cite{beldi}              & \textcolor{green!50!black}{Yes}  & \textcolor{green!50!black}{Yes}       & \textcolor{green!50!black}{Yes}           & \textcolor{green!50!black}{Yes}  & \textcolor{red!70!black}{No} \\
      Boki~\cite{boki}               & \textcolor{green!50!black}{Yes}  & \textcolor{green!50!black}{Yes}       & \textcolor{green!50!black}{Yes}           & \textcolor{green!50!black}{Yes}  & {\color[HTML]{583200}Caching} \\
      \midrule
      \textbf{\sn} & \textcolor{green!50!black}{Yes}  & \textcolor{green!50!black}{Yes}       & \textcolor{green!50!black}{Yes}           & \textcolor{green!50!black}{Yes} & \textcolor{green!50!black}{Co-location} \\
      \bottomrule
  \end{tabular}

  } 
  \caption{\sn provides similar or stronger guarantees than comparable platforms while improving performance by co-locating compute and data. CC means causal consistency; TCC means transactional causal consistency.}
  \label{tab:semantics-comparison}
\end{table}
\section{Fault-Tolerant Workflows}
\label{sec:execution}

We now describe how we leverage \sn's integration of functions and data to efficiently implement the workflow fault-tolerance guarantees defined in \S\ref{sec:semantics}: run-to-completion workflow execution and exactly-once function execution.

\subsection{Handling Machine Failures}
\label{sec:exactly_once_semantics}

\sn must enforce its execution guarantees despite failures of frontend dispatchers or DBMS backend servers.
Most distributed DBMSs can recover from failures of their servers, typically using replication and logging.
We assume that if any single server fails, the DBMS can recover without loss of availability by failing over to a replica, so workflow execution is unaffected.
If multiple servers fail, the DBMS can recover without data loss from durable logs, so dispatchers must wait until recovery is complete, then resume execution.

If a dispatcher fails during workflow execution, clients with a pending workflow invocation time out, then resubmit their invocation to resume the partially executed workflow on a new dispatcher.
To uniquely identify workflows for resumption, clients generate a unique ID for each workflow invocation and prefix it with a database-generated unique client ID.
The new dispatcher must resume workflow execution from where the failed one left off, finishing the workflow without re-executing any functions that have already been completed.
Because \sn runs functions transactionally as stored procedures, we can make this possible by instrumenting functions to transactionally \emph{record} their outputs (serialized in a binary format) in the DBMS before returning.
Each recorded output is associated with a unique function invocation ID, derived from the workflow ID, and is retained only for the lifetime of the workflow.
During retry, the dispatcher resumes a workflow from the beginning and (re-)dispatches each function.
Functions first check for a record from the earlier execution and, if they find one, return it instead of executing.

A limitation of this current implementation is that it relies on clients to detect dispatcher failures.
In future work, we plan to fix this by having dispatchers write ahead workflow metadata to the DBMS and ping each other in a decentralized manner to detect failures (using the DBMS for discovery).
Then, a dispatcher which detects another's failure could retrieve its pending workflows from the DBMS and complete their execution without client involvement.

\subsection{Optimizing Function Recording}
\label{sec:exactly_once_semantics_alg}

The protocol described in the previous section records every function's output in the database, but doing this naively incurs overhead of up to 2.2$\times$ (as we show in \S\ref{sec:eval-overhead-analysis}) because it requires performing additional database lookups and updates in each function.
However, we can reduce this overhead to <5\% (across all workloads we tested) by recognizing that some functions can be safely re-executed without violating exactly-once semantics, so their outputs need not be recorded.
For example, if an entire workflow is read-only,
it can be safely re-executed if its original execution failed,
so we need not record any of its functions.
Therefore, we develop a new algorithm, called \emph{selective function recording} (SFR), to determine, using static analysis when a workflow is registered, which functions must be recorded and which can be safely re-executed.

We must record any function performing a DBMS write to ensure writes are not re-executed (re-executing external calls is fine as they must be idempotent).
Moreover, we must record a read-only function if there exist disjoint paths from it to multiple different recorded functions,
or to at least one recorded function and the sink.
This guarantees that two recorded functions which depend on a value computed by an ancestor will 
always observe the same value from that ancestor,
even if workflow execution is restarted.
To determine whether disjoint paths exist, we search for all recorded functions (or the sink) reachable without traversing another recorded function; if there are more than one of these, there are disjoint paths, and the function must be recorded.

\begin{figure}[t!]
	\includegraphics[width=\linewidth]{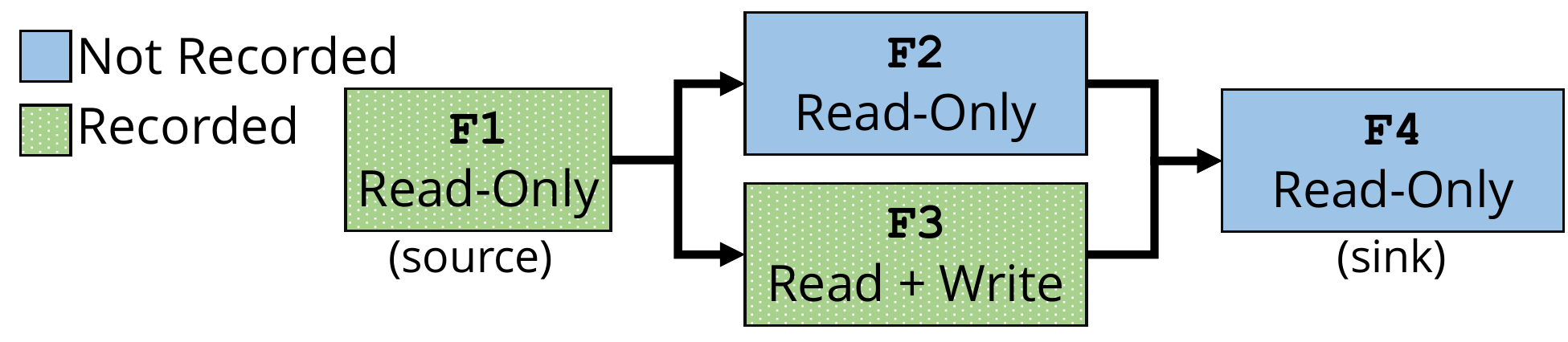}
	\caption{
		Example of \sn's selective recording algorithm \algoname. \texttt{F3} is recorded as it performs a write. \texttt{F1} must also be recorded to avoid inconsistent outputs to \texttt{F2} and \texttt{F3}.
	}
	\label{fig:fault_tolerance_logging}
\end{figure}

We sketch \algoname in Algorithm~\ref{alg:fault_tolerance} and
provide an example in Figure~\ref{fig:fault_tolerance_logging}.
\texttt{F3} is recorded for performing writes, but \texttt{F1} is also recorded despite being read-only.
Suppose \texttt{F1} was not recorded and a dispatcher crashed after executing \texttt{F1} and \texttt{F3}.
Upon re-execution, \texttt{F1} may return a different value than it originally did because some data was changed by an unrelated function.
If that happened, \texttt{F2} would return an output based on the new output of \texttt{F1}, but \texttt{F3} would return its recorded output based on the original output of \texttt{F1}.
This causes an inconsistency that violates exactly-once semantics, so we must record \texttt{F1} to prevent it.

\begin{algorithm}[t!]
	\small
    \begin{algorithmic}[1]
    \Function{\algoname}{$W$}  \Comment Input $W$: the workflow graph.
    \State \{$f_1$,...,$f_n$\} = topoSort($W$) \Comment $f_1$ is source, $f_n$ is sink.
    \State $Recorded = \{\}$
    \For{$f_i \in \{f_n...f_1\}$}    \Comment Traverse from sink back to source.
    \If{hasWrite($f_i$)}
      \State $Recorded$.add($f_i$)
    \Else
      \Statex \LeftComment{3}{BFS search all recorded functions (or the sink) }
      \Statex \LeftComment{3}{reachable without traversing a recorded function.}
      \State $RF = BFSFindReachable(f_i, Recorded \cup \{f_n\}$)
      \If{$RF$.size() > 1}
        \State $Recorded$.add($f_i$)
      \EndIf
    \EndIf
    \EndFor
    \State return $Recorded$
    \EndFunction
  
    \end{algorithmic}
    \caption{\small \algoname: Selective Function Recording}
    \label{alg:fault_tolerance} 
\end{algorithm} 

\noindent\textbf{Correctness.} \algoname identifies a set of recorded functions such that if a workflow is resumed following the protocol described in the previous section (re-executing non-recorded functions, returning recorded outputs of recorded functions), the effect of the workflow on application state is the same as if every function in the workflow were executed exactly once.
Recording extra functions cannot provide stronger guarantees because in the absence of disjoint paths to different recorded functions, a non-recorded function must have a single recorded function descendant that is the ancestor of all other recorded function (or sink) descendants and can provide a consistent output during failure recovery.
We do not guarantee we find the minimal set of recorded functions as semantic information about functions may obviate the need to record a function (for example, if we knew a function returned a constant to another function, we could ignore that edge for this algorithm), but this is outside the scope of this paper.

\noindent\textbf{Complexity.}
Because \algoname traverses a workflow from the sink back to the source, we can memoize workflow graph search, so we only need to traverse each workflow graph edge once.
Therefore, the time complexity of \algoname is $O(V + E)$ where $V$ is the number of functions, and $E$ is the number of edges in the workflow graph.
We only run this algorithm once per workflow when the workflow is registered.

\subsection{Handling Function Failures}

\sn must additionally enforce its execution guarantees despite failures or errors in individual functions.
If a function fails due to recoverable or transient errors (e.g., the failure of a DB server that can fail over to a replica), the dispatcher retries it.
However, if an error is unrecoverable (e.g., a constraint violation or function runtime exception), there is no choice but for the DBMS to abort and roll back its containing transaction.
From there, \sn continues workflow execution, but propagates a failure notification to any downstream functions which had as inputs the output of the failed function.
This enforces run-to-completion workflow execution while giving developers control over how their workflows handle failures.

\section{Observability}
\label{sec:provenance}

Developers often require information on how applications interact with data for debugging, monitoring, and auditing use cases,
for example to verify a program did not improperly access private data.
In existing FaaS platforms, this information is fundamentally difficult
to collect because platforms lack visibility into how functions manage data,
so developers must perform extensive manual logging across many short-lived functions.
In this section, we discuss how we leverage \sn's tight integration with the database to build a tracing layer that
automatically records these interactions with minimal cost. 

\subsection{Observability Interface}

\sn instruments workflows to trace the history of workflow and function executions,
instruments queries to log database operations,
then combines this information to create a complete record of application interactions with data. 
Specifically, \sn records for each data item all function executions that accessed or modified it.
The tracing layer automatically spools this information to an analytical database
(in our implementation, Vertica~\cite{vertica}) for long-term storage and analysis.
We use a separate analytical database because it is better optimized for large observability queries than a transactional DBMS.
Storage policies such as data retention rules are implemented by this database.
Within the analytical database, information is organized into tables.
For captured workflow information, \sn creates a function invocations table per application:
\begin{tcolorbox}[left=0pt,right=0pt,top=0pt,bottom=0pt,before skip=5pt,after skip=5pt]
	\begin{Verbatim}[fontfamily=lmtt,fontsize=\small]
FunctionInvocations (func_id, timestamp,
      function_name, workflow_name, workflow_id)
	\end{Verbatim}
\end{tcolorbox}
\noindent \texttt{func\_id}, the primary key, is a unique ID per function invocation.  \texttt{workflow\_id} is a unique ID per workflow invocation.  Both are defined in \S\ref{sec:exactly_once_semantics}.

For each table used by an application, \sn creates an event table
for captured operations on that table:
\begin{tcolorbox}[left=0pt,right=0pt,top=0pt,bottom=0pt,before skip=5pt,after skip=5pt]
	\begin{Verbatim}[fontfamily=lmtt,fontsize=\small]
TableEvents (func_id, timestamp, event_type, 
      query, [record_data...])
	\end{Verbatim}
\end{tcolorbox}
\noindent \texttt{event\_type} can be \texttt{insert, delete, update}, or \texttt{read}; \texttt{query} is the query string; \texttt{func\_id} is a foreign key referencing \texttt{FunctionInvocations}.

The information stored in these tables enables efficient execution of useful observability queries for
debugging, monitoring, and auditing FaaS applications.
For example, developers could query if a function improperly accessed private data and what it did with the data.
We evaluate case studies in \S\ref{sec:case_studies}.

\subsection{Implementing Tracing Layer}
\label{sec:provenance_capture}
We leverage \sn's tight integration with data to adapt database techniques like change data capture and query rewrites~\cite{glavic2009perm,arab2014generic} to a FaaS setting,
building a tracing layer to capture application interactions with data efficiently.
When a function executes, the tracing layer adds an entry to \texttt{FunctionInvocations}.
When a function performs a database operation, it automatically records metadata such
as the function ID to \texttt{TableEvents}.
For write operations, it also records updated data.
For read operations, the tracing layer modifies the query to return the primary keys of all retrieved rows
(in addition to the requested information), then records them to identify each accessed record.
To minimize overhead, it only captures rows that are actually retrieved, not rows that are accessed incidentally (e.g., by an aggregation that may access thousands of rows).
The tracing layer also logs the queries themselves so this additional read information may be reconstructed later if it is needed for an investigation.

The challenge in capturing information on database operations is performing it efficiently
while providing guarantees about what information is captured.
To capture writes, the tracing layer relies on DBMS change data capture
to transactionally export information.
However, read capture is more difficult, both because existing DBMSs do not have built-in read capture capability 
and because reads are numerous so overhead may be higher.
Reads are captured by instrumenting the \texttt{execQuery} function (Figure~\ref{fig:dbos_interface}).
The tracing layer maintains a circular buffer inside each DBMS server's memory.
Whenever a read occurs in \texttt{execQuery}, the tracing layer appends its information to this buffer.
Periodically, it flushes the buffer to the remote analytical database.
Tracing is robust to function failures and re-executions because re-executions are recognized and deduplicated using their shared IDs.
However, captured read information may be lost if
the database server crashes while information is in the buffer;
if developers cannot tolerate data loss, we can optionally place the buffer on disk
to eliminate loss at some performance cost.
\section{Implementation}
\label{sec:implementation}

We implement \sn's tightly integrated architecture (\S\ref{sec:system_overview}) by wrapping a distributed DBMS and compiling functions to DBMS stored procedures.

\subsection{Choosing a DBMS}
\sn requires a distributed DBMS with four properties: 

\begin{itemize}
	\item Supports ACID transactions.
	\item Supports running user code in a non-SQL language transactionally in stored procedures.
	\item Supports change data capture (for observability, \S\ref{sec:provenance_capture}).
	\item Supports elastic DBMS cluster resizing.
\end{itemize}

\noindent While many DBMSs have these properties (e.g., SingleStore~\cite{singlestore}, Yugabyte~\cite{yugabytedb}), we chose VoltDB as it could most efficiently execute our target workloads.
Most distributed DBMSs, including VoltDB, scale by partitioning data.
We observe that in our target workloads, almost all transactions are single-sited~\cite{hstore} and access data in only a single partition.
VoltDB executes these transactions efficiently, running them to completion in memory without needing locks.
However, a limitation of VoltDB is that it is less efficient at multi-sited transactions; a transaction must hold a global lock to access data on multiple partitions.
There has been recent research on addressing this~\cite{lotus}, but we leave the efficient implementation of multi-sited transactions to future work.

\subsection{Compilation}
\label{sec:compilation}

When developers register functions and workflows in \sn, it compiles each function to a stored procedure, a routine in a non-SQL language that runs natively as a DBMS transaction.
In our implementation, functions provide the same guarantees as VoltDB transactions: they are ACID and serializable.
To implement multi-function transactions, \sn compiles all involved functions to a single stored procedure.
Compilation happens in two steps.
First, \sn instruments each function to capture application-database interactions for observability (\S\ref{sec:provenance}) and to record its execution for exactly-once semantics (\S\ref{sec:execution}).
Then, \sn compiles the instrumented function (or multi-function transaction) into a stored procedure and registers it in the DBMS.

\sn extends the DBMS stored procedure interface, so it can compile any function that uses its programming interface (Figure~\ref{fig:dbos_interface}) and follows the rules outlined in \S\ref{sec:programming_model}.
Additionally, in our VoltDB-based implementation, because VoltDB can efficiently execute single-sited transactions,
we let developers specify if a function (or multi-function transaction) is single-sited and, if so,
which function input specifies the site.

\section{Evaluation}
\label{sec:evaluation}

\begin{figure}[t!]
	\includegraphics[width=\linewidth]{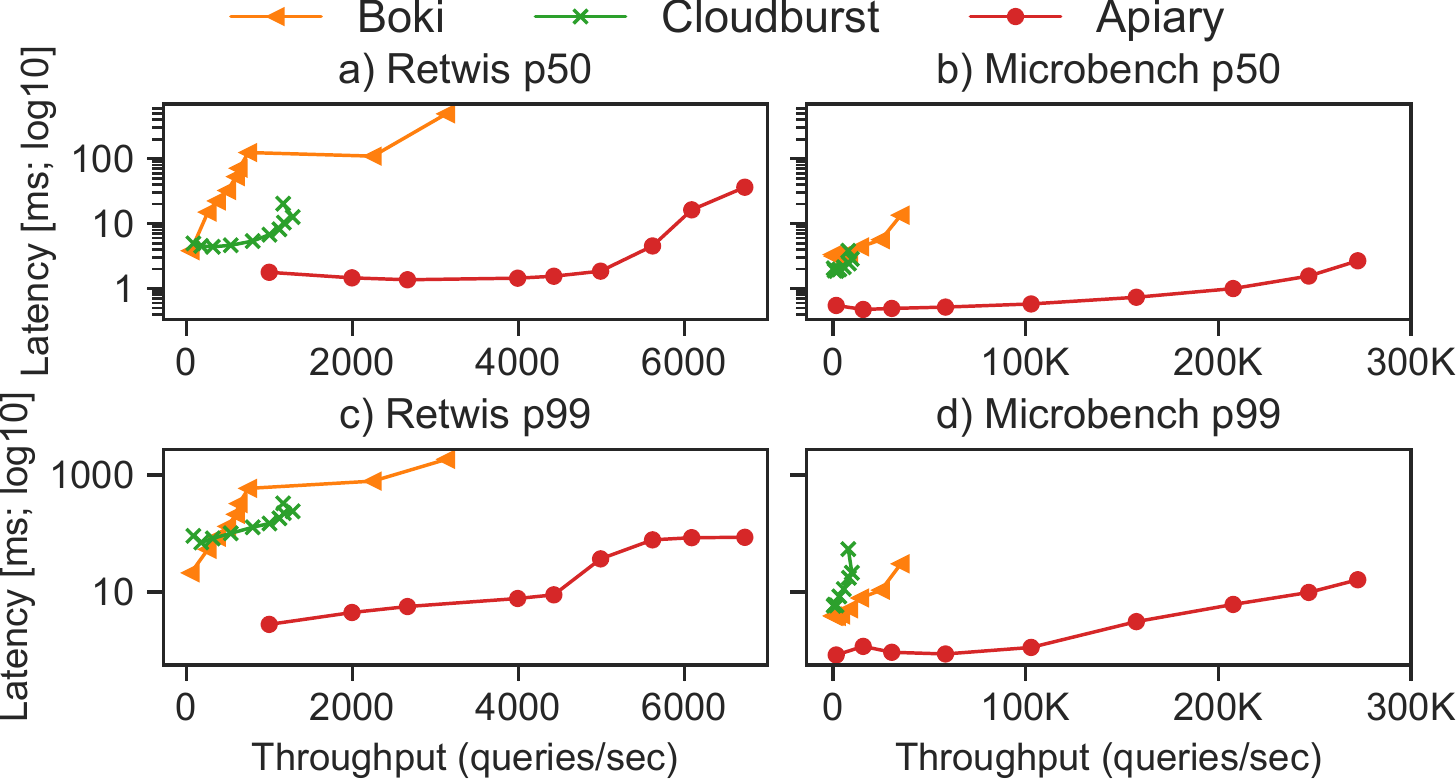}
	\caption{
		Throughput versus latency for \sn, Boki, and Cloudburst on Retwis and on a microbenchmark.
	}
	\label{fig:boki_cloudburst_comparison}
\end{figure}

We evaluate \sn with widely-used microservice and web serving workloads as well as microbenchmarks, showing that:

\begin{enumerate}
    \item By physically co-locating compute and data,
    \sn outperforms production FaaS systems by 7--68$\times$ and research systems by 2--27$\times$ (Figures~\ref{fig:baseline_comparison}, ~\ref{fig:boki_cloudburst_comparison}).
    \item By selectively instrumenting functions using the \algoname algorithm, \sn provides fault tolerance with overhead of <5\% compared to 2.2$\times$ for a naive solution (Figure~\ref{fig:fault_tolerance_analysis}).
    \item By instrumenting database operations and functions,
    \sn captures information on application-database interactions critical to observability with overhead of <15\% as compared to 92\% with manual logging (Figure~\ref{fig:provenance_analysis}).
\end{enumerate}

\subsection{Experimental Setup}

We implement \sn in \textasciitilde10K lines of Java code, which we will open source upon paper acceptance.
We use VoltDB~\cite{voltdb} v9.3.2 as our DBMS backend and Vertica~\cite{vertica} v10.1.1 for analytics data.
For communication between clients and frontend servers,
we use JeroMQ~\cite{jeromq} v0.5.2 over TCP.

In all experiments where not otherwise noted, we run on Google Cloud using \texttt{c2-standard-8} VM instances with 8 vCPUs and 32GB DRAM.
We use as a DBMS backend a cluster of 40 VoltDB servers with 8 VoltDB partitions per VM.
For high availability, we replicate each partition once, as is common in production.
For fairness, we use the same VoltDB cluster as the storage backend for our baselines.
To ensure we can fully saturate this DBMS backend, we run 45 \sn frontend VMs and generate requests using 15 remote client VMs.
Each client VM runs on a \texttt{c2-standard-60} instance with 60 vCPUs and 240GB DRAM.
We spool observability data to a cluster of 10 Vertica servers running on a separate set of VMs from the VoltDB cluster.
All experiments run for 300 seconds after a 5-second warmup.

\subsection{Baselines}

We compare \sn to four baselines, ranging from production platforms to the latest research systems.

\noindent\textbf{OpenWhisk.}  OpenWhisk (OW)~\cite{openwhisk} is a popular open-source production FaaS platform. 
We implement each of our workloads in the OW Java runtime, performing all business logic in an OW function
but storing and querying data in an external VoltDB cluster.
We coordinated with OW developers to tune our OW setup.
Since OW cannot efficiently run workflows, we implement each workload in a single OW function, eliminating communication between functions.
Additionally, we pre-warm OW function containers and only measure warm-start performance.
In our experiments, we use 45 \texttt{c2-standard-8} VMs as OW workers. To maximize OW performance, we use 5 controllers, each on a \texttt{c2-standard-60} instance that manages 9 workers, load balancing between sub-clusters.

\noindent\textbf{RPC Servers.}  Most microservices today are deployed in long-running RPC servers with separate application and DBMS server machines~\cite{deathstarbench,laigner2021data}.
We implement each of our workloads this way, running all business logic in RPC servers
but storing data in an external VoltDB cluster and accessing it using VoltDB stored procedures.
For fairness, we use the same communication library as \sn (JeroMQ, chosen because we found it outperforms alternatives like gRPC), re-implement each microservice in Java following its original architecture,
and use long-lived connections with the DBMS.
In our experiments, we use a setup identical to \sn but with all frontend servers replaced with RPC servers.

\noindent\textbf{Boki.}  Boki~\cite{boki} is a recent research system supporting transactional FaaS.
We use Boki as a baseline because it is representative of a class of transactional FaaS systems (discussed in \S\ref{sec:semantics-comparison}),
but is additionally co-designed with in-memory local caches for high performance.
We use the experimental setup described in the Boki paper, deploying 8 storage nodes, 3 sequences, and 8 workers,
each on an AWS EC2 \texttt{c5d.2xlarge} instance with 8 vCPUs and 16GB DRAM.
We coordinated with the Boki authors to tune our setup.

\noindent\textbf{Cloudburst.} Cloudburst~\cite{cloudburst} is a recent research system for stateful FaaS that provides causal consistency.
We use Cloudburst as a baseline because it is an influential system that represents a class of FaaS platforms which are built around causal consistency and improve performance with in-memory local caches (discussed in \S\ref{sec:semantics-comparison}).
To maximize Cloudburst performance and ensure consistency of experimental results,
we disabled its autoscaler and manually pinned function executors to every available worker thread.
We otherwise run Cloudburst unmodified in its most-performant last-writer wins mode.
Similar to the Boki setup and following recommendations from the Cloudburst authors,
we deploy 1 Anna KVS node, 4 scheduler nodes, and 8 worker nodes,
each on an AWS EC2 \texttt{c5.2xlarge} instance with 8 vCPUs and 16GB DRAM.

For fairness, when comparing \sn to Boki and Cloudburst, we use 8 VoltDB servers and 8 frontend servers.

\subsection{Microservice Workloads}

We evaluate \sn using three microservice benchmarks,
each commonly used in previous microservices and FaaS papers.
As shown in Table~\ref{tab:workload_baseline}, these workloads cover a large design space for data-centric FaaS applications.

\noindent\textbf{Shop.} This benchmark, adapted from a Google Cloud demo~\cite{hipster_shop}, simulates a service
where users browse an online store, update their shopping cart, and check out items.

\noindent\textbf{Hotel.}  This benchmark, from DeathStarBench~\cite{deathstarbench},
simulates searching and reserving hotel rooms.  Our implementation contains a multi-function transaction similar to Figure~\ref{fig:example-code}, where validation and reservation are performed transactionally.

\noindent\textbf{Retwis.}  This benchmark, from Redis~\cite{retwis}, simulates a Twitter-like social network,
where users follow other users, make posts, and read a ``timeline'' of the most recent posts
of all users they follow.
We use the same Retwis parameters as Cloudburst~\cite{cloudburst}: we create 1000 users, each following 50 other users, and pre-load 5000 posts.

\begin{table}[t]
\resizebox{\linewidth}{!}{%
\centering
\small
\setlength{\tabcolsep}{3pt}
\begin{tabular}{@{}cc|cccccc@{}}
\toprule
 & {\color[HTML]{333333}} &  &  &  &  & & \\
\multirow{-2}{*}{\textbf{Workload}} & \multirow{-2}{*}{\textbf{Operation}} & \multirow{-2}{*}{\textbf{Ratio}} & \multirow{-2}{*}{\begin{tabular}[c]{@{}c@{}}\textbf{Read-}\\\textbf{Only}?\end{tabular}} & \multirow{-2}{*}{\begin{tabular}[c]{@{}c@{}}\textbf{Access}\\ \textbf{Rows}\end{tabular}} & \multirow{-2}{*}{\begin{tabular}[c]{@{}c@{}}\textbf{RPCs for}\\\textbf{$\mu$Services}\end{tabular}} & \multirow{-2}{*}{\begin{tabular}[c]{@{}c@{}}\textbf{ \# of}\\ \textbf{Txns.}\end{tabular}} & \multirow{-2}{*}{\begin{tabular}[c]{@{}c@{}}\textbf{\# of SQL} \\ \textbf{Queries}\end{tabular}}  \\\midrule
 & Browsing & 80\% & Yes & 8 & 2 & 1 & 1\\
 & CartUpdate & 10\% & No & 1 & 2 & 1 & 2\\
\multirow{-3}{*}{Shop} & Checkout & 10\% & No & 5 & 6 & 3 & 5\\\midrule
 & Search & 60\% & Yes & 30 & 4 & 6 & 22\\
 & Recommend & 39\% & Yes & 1 & 2 & 1 & 1\\
\multirow{-3}{*}{Hotel} & Reservation & 1\% & No & 5 & 2 & 2 & 5\\\midrule
 & GetTimeline & 90\% & Yes & 550 & 3 & 51 & 51\\
\multirow{-2}{*}{Retwis} & Post & 10\% & No & 1 & 2 & 1 & 1\\\bottomrule
\end{tabular}
}
\caption{Microservice benchmark information.
RPCs are for the RPC Servers baseline; \sn and OW only require one client-server RPC.
\sn only requires one DB round trip per transaction, but the baselines require one per SQL query.}
\label{tab:workload_baseline}
\end{table}

\subsection{End-to-End Benchmarks}
\label{sec:eval-microservices}

\begin{figure}[t!]
	\includegraphics[width=\linewidth]{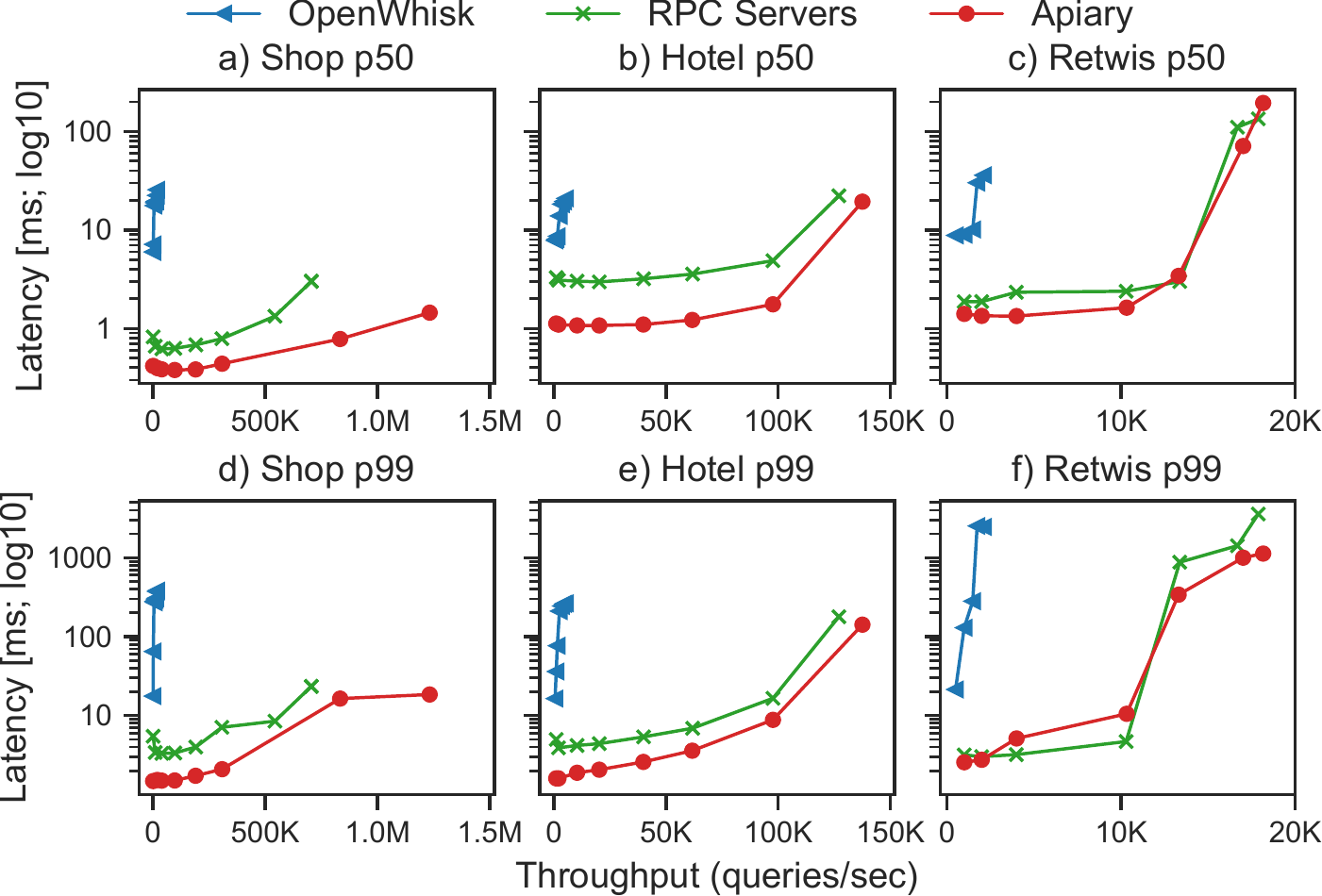}
	\caption{
		Throughput versus latency for \sn and the OpenWhisk (OW) and RPC Servers baselines
		on all benchmarks.
	}
	\label{fig:baseline_comparison}
\end{figure} 

We first compare \sn performance with the OW and RPC Servers baselines on our three microservice workloads,
showing results in Figure~\ref{fig:baseline_comparison}.
For each benchmark, we vary offered load (sent asynchronously following a uniform distribution)
and observe throughput and latency.
For all three workloads, maximum throughput achieved by \sn is greater for Shop (1.2M RPS) than Hotel (144K RPS)
and for Hotel than Retwis (20K RPS).
This is because most Shop operations access a single customer's cart,
while most Hotel operations look up data for several hotels and most Retwis
operations access data for several dozen users (``Access Rows'' in Table~\ref{tab:workload_baseline}).

We find that \sn significantly outperforms the RPC Servers baseline on two benchmarks and performs on par on the third -- even though \sn offers more features (like observability information capture)
and stronger guarantees (like ACID functions and fault-tolerant workflows).
\sn outperforms the RPC Servers baseline due to reduced communication overhead:
because it compiles services to stored procedures that run in the database server, it requires fewer round trips to perform database operations (Table~\ref{tab:workload_baseline}).
\sn achieves 1.6--3.4$\times$ better median and tail latency than RPC servers on Shop and Hotel,
where each transaction executes many database queries which each require an RTT in the baseline but not in \sn.
It matches the baseline latency for Retwis, where each transaction executes only a single query.
\sn and RPC servers achieve similar maximum throughput for Hotel and Retwis, where throughput is bottlenecked by the database,
but \sn improves throughput by 1.75$\times$ for Shop, where the bottleneck is communication.

\sn dramatically outperforms OW on all three benchmarks.
Due to a combination of scheduling, container initialization, message passing, and communication overhead 
(analyzed in \S\ref{sec:eval-microbench}),
\sn improves throughput by 7--68$\times$ and median and tail latency by 5--14$\times$ compared to OW.

These results establish that,
for our target applications, separating function execution from data management is inefficient.
A conventional FaaS platform not only requires the same number of storage servers as \sn to host and manage data,
but also needs compute workers to run application logic.
However, because application logic is computationally bottlenecked by database operations
(as shown in Figure~\ref{fig:exec_time_breakdown}),
these compute workers contribute little but add significant communication overhead.
Thus, \sn's architecture reduces communication overhead, uses resources more efficiently,
and, as we will show in Section~\ref{sec:cost_analysis}, reduces the cost of deployment.

\begin{figure}[t!]
	\centering
	\includegraphics[width=0.95\linewidth]{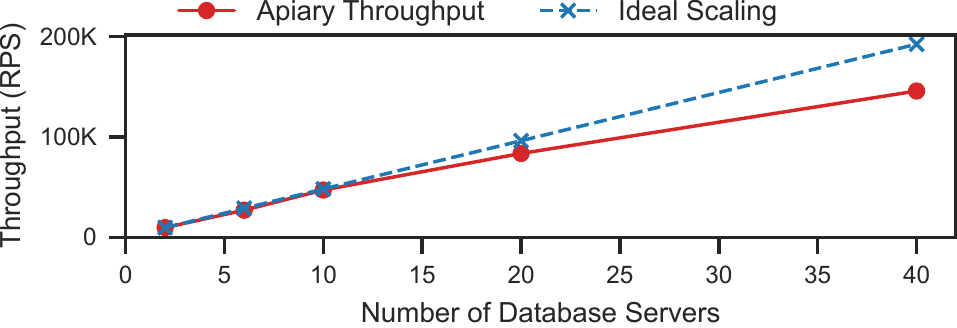}
	\caption{
		Maximum throughput for \sn on Hotel with a varying number of database servers.
		We extrapolate ``ideal scaling'' linearly from a single replicated server (two servers total).
	}
	\label{fig:dbos_scalability}
\end{figure}

\noindent\textbf{Scalability.}
We also evaluate the scalability of \sn,
measuring the maximum throughput \sn can achieve with varying numbers of database servers.
We show results for the Hotel benchmark in Figure~\ref{fig:dbos_scalability}, but obtained similar results for Shop and Retwis.
We measure from 2 to 40 database servers (16 to 320 data partitions), 
beginning with 2 servers because each server needs a replica.
We find that \sn scales well; with larger numbers of servers, performance was mainly limited by VoltDB's overhead of managing a large network mesh.

\subsection{OpenWhisk Performance Analysis}
\label{sec:eval-microbench}

\begin{figure}[t!]
	\includegraphics[width=\linewidth]{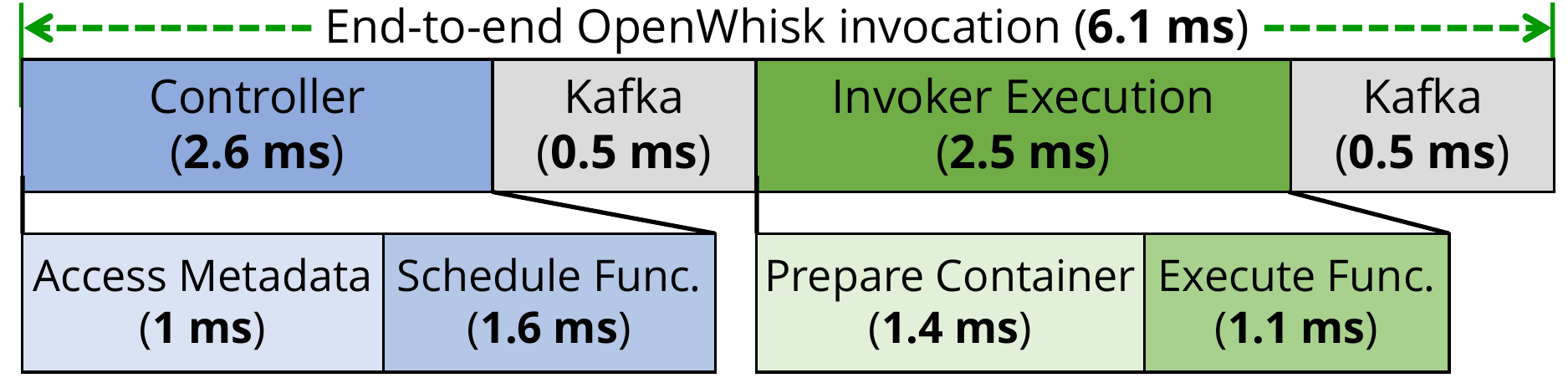}
	\caption{
		Latency breakdown for an OpenWhisk function invocation performing a point database update.
	}
	\label{fig:ow_latency_breakdown_details}
\end{figure} 

To further investigate the performance difference between \sn and production FaaS systems like OW,
we analyze OW performance on a microbenchmark of a single OW function that retrieves and increments a counter stored in VoltDB.
We invoke this function 100K times and measure the average latency of each step.

As we show in Figure~\ref{fig:ow_latency_breakdown_details}, OW adds significant overhead to a function invocation.
Each invocation is first handled by a controller which performs bookkeeping operations using function metadata stored in CouchDB (1 ms) before scheduling the invocation to an invoker/worker node (1.6 ms).
OW uses Apache Kafka for controller-invoker communication, incurring 1 ms of round-trip latency.
Once the invoker receives a request, it resumes the execution of an already-warm container (1.4 ms).
The function then executes in 1.1 ms.
We emphasize that this high overhead is not unique to OW;
other popular production FaaS systems have similar architecture and performance characteristics.
\sn avoids this overhead because it integrates function execution and data management and stores all state in the backend DBMS, reducing communication overhead and avoiding external state management.

\subsection{Comparing with Boki and Cloudburst}
\label{sec:boki-cloudburst-comparison}

We next compare \sn performance with Boki and Cloudburst.
We use the Retwis benchmark because both Cloudburst and Boki use it in their evaluation and provide open-source implementations of it, so we can be sure our comparison is fair.
Retwis is read-heavy, so to evaluate the performance impact of writes,
we use a microbenchmark which retrieves and increments a counter associated with a key.
We use 80K counters to minimize aborts due to write-write conflicts in Boki
while still ensuring all counters fit into in-memory cache.

Looking first at Boki, we find that \sn improves throughput by 2.1$\times$ on Retwis
and 10.1$\times$ on the microbenchmark.
We see similar improvements for median and tail latencies.
Unlike \sn, Boki must frequently perform non-local reads in the presence of writes
to update its caches and enforce its snapshot isolation guarantee.
Thus, we expect Boki to perform relatively better on read-heavy Retwis
and relatively worse on the write-heavy microbenchmark, where most reads are non-local.
Our experiments confirm this hypothesis.

Looking next at Cloudburst,
we find that \sn improves throughput by 5.2$\times$ on Retwis and 27.7$\times$ on the microbenchmark, with similar trends for latencies.
The performance difference is surprising because both \sn and Cloudburst perform all data access and updates locally,
using stored procedures in \sn and asynchronously-synchronized local caches in Cloudburst (though \sn provides stronger guarantees than Cloudburst: ACID transactions versus causal consistency).
Digging deeper, we find the performance difference comes largely from the more efficient implementation of \sn:
a read from a local cache in Cloudburst takes 300~\textmu s (this is high because Cloudburst is implemented in Python) as compared to <20~\textmu s for a VoltDB read in \sn
(this adds up because Retwis contains several reads and Cloudburst does not batch them)
and additionally Cloudburst incurs 2.2~ms of executor and scheduler overhead for each function execution
as compared to 300~\textmu s in \sn.

\subsection{Fault-Tolerant Workflows Performance Analysis}
\label{sec:eval-overhead-analysis}

\begin{figure}[t!]
	\includegraphics[width=\linewidth]{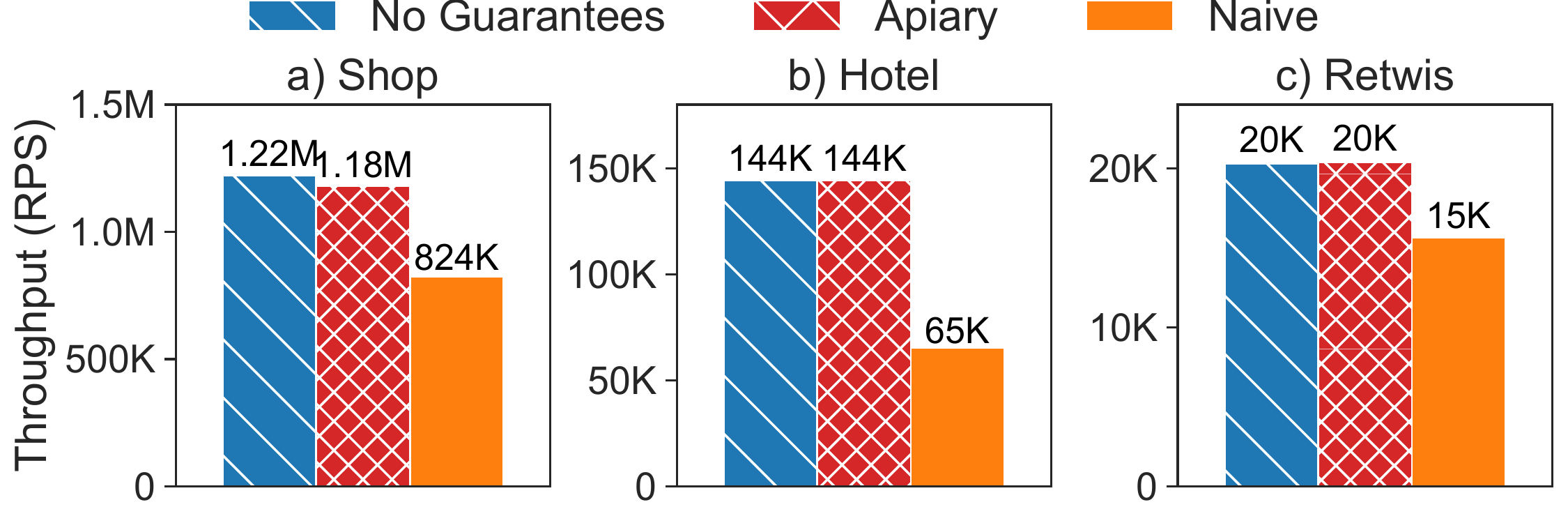}
	\caption{
		Maximum achievable throughput for \sn without workflow fault-tolerance guarantees,with the guarantees, and with a naive implementation of the guarantees.
	}
	\label{fig:fault_tolerance_analysis}
\end{figure}

We now analyze the performance impact of \sn's workflow fault-tolerance guarantees.
\sn uses \algoname(\S\ref{sec:exactly_once_semantics}) to selectively record function outputs in the DBMS to avoid re-executing them when resuming a failed workflow.
We evaluate the overhead of our guarantee and compare it to a more naive implementation (similar to prior work~\cite{beldi}) that records all function executions, showing results in Figure~\ref{fig:fault_tolerance_analysis}.
We find our guarantee incurs overhead of <5\%, but this low overhead is only possible because \algoname lets us record selectively: only 25\% of Shop, 0.25\% of Hotel, and 0.2\% of Retwis transaction executions must be recorded.
By contrast, the naive implementation reduces throughput by 1.3--2.2$\times$.

\subsection{Enhancing Observability with \sn}
\label{sec:case_studies}

\noindent\textbf{Data Tracing Performance Analysis.}
To analyze the performance impact of \sn's observability tracing (\S\ref{sec:provenance}), we measure \sn performance with and without capture.
We also measure the performance of a ``manual logging'' baseline that represents how observability information is captured
by application developers utilizing existing FaaS platforms: by manually logging to local disk files that are later exported by monitoring software like AWS Cloudwatch.

\begin{figure}[t!]
	\includegraphics[width=\linewidth]{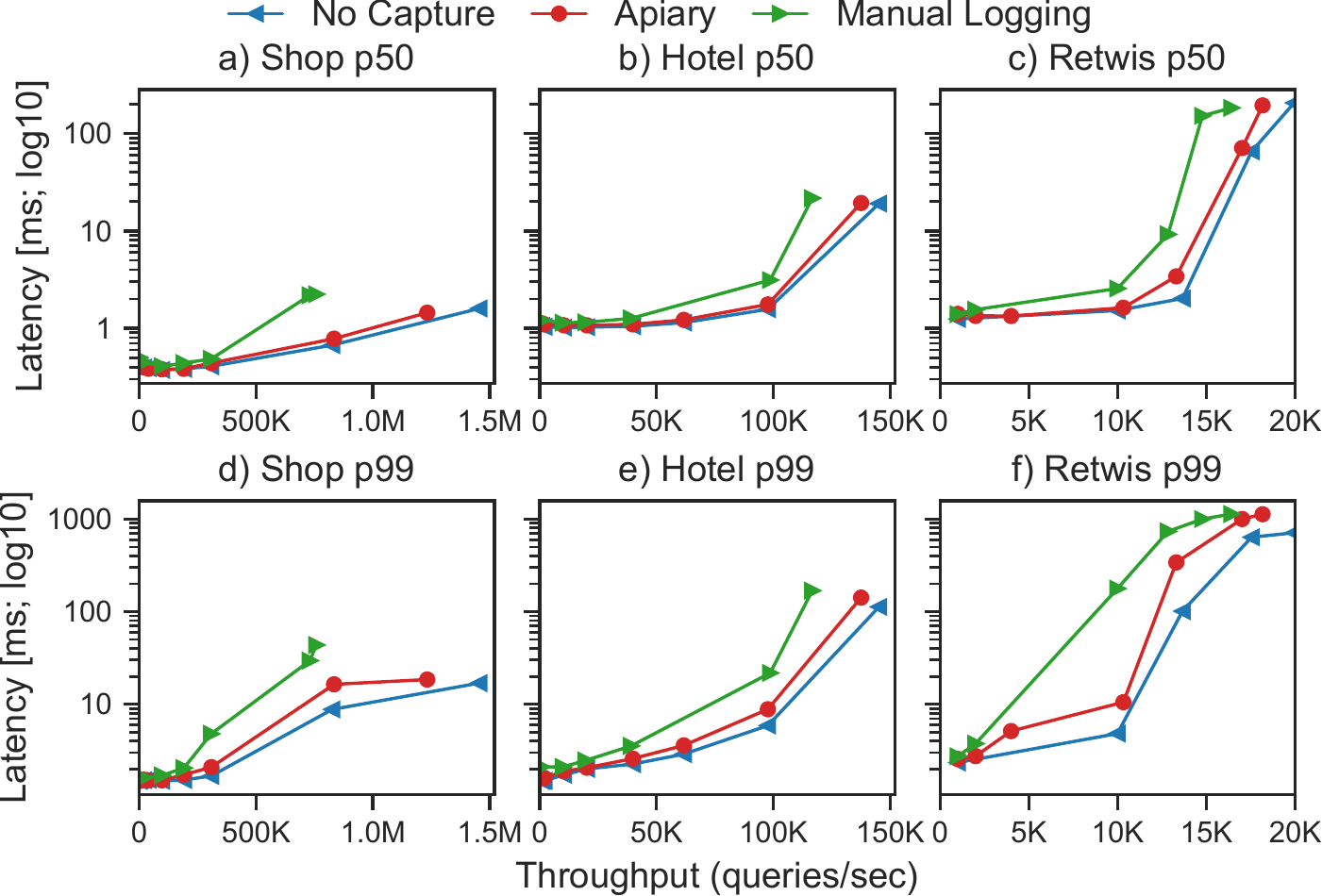}
	\caption{
		Throughput versus latency for \sn, no observability capture,
		and a manual logging baseline
		on all benchmarks.
	}
	\label{fig:provenance_analysis}
\end{figure}

We show results for all workloads in Figure~\ref{fig:provenance_analysis}.
At low load, both \sn and manual logging slightly increase
latency (both median and tail) by up to 10\%.
At high load, \sn adds throughput overhead of up to 15\% while manual logging adds overhead of up to 92\%.
\sn data tracing overhead is low because we minimize the cost on the critical path,
buffering captured observability data in the database's memory and asynchronously exporting it in large batches.

\noindent\textbf{Case Studies.}
We next evaluate the value and practicality of \sn's data tracing.
We execute 150M Shop operations, generating 1.2B rows of traced data,
and export this data to a single Vertica server, finding it compresses to just 12.4GB of disk space.
We then use this dataset to show how \sn can handle queries from tasks of interest to our industrial partners. 

\noindent\textbf{Debugging.}  Our first query is ``What was the state of some record X when it was read by this particular function execution?''
This query might be used to determine what input caused a function abort.
\sn can answer this query because it records in \texttt{TableEvents} all changes to data, so we can retrieve the last update to record X before the problematic function execution began.
For instance, we can use a single SQL query to find the state of record X at a time TS:
\begin{tcolorbox}[left=0pt,right=0pt,top=0pt,bottom=0pt,before skip=5pt,after skip=5pt]
	\begin{Verbatim}[fontfamily=lmtt,fontsize=\footnotesize]
SELECT reco_data FROM TableEvents
WHERE event_type IN ('insert', 'update') AND rec_id=X
  AND timestamp <= TS ORDER BY timestamp DESC LIMIT 1; 
	\end{Verbatim}
\end{tcolorbox}
We execute this query on the largest event table in our Shop dataset (840M rows and 7GB storage) to find the exact state of a record when it was retrieved by a particular Shop execution; the average query latency across five runs is 4.3 seconds.

\noindent\textbf{Downstream Provenance.} Our second query is ``Find all records updated by a workflow that earlier read record X.''
This query is useful for taint tracking, for example if record X contains misplaced sensitive information.
We can answer this query by scanning for functions that read record X,
then returning the write sets of later functions in their workflows:
\begin{tcolorbox}[left=0pt,right=0pt,top=0pt,bottom=0pt,before skip=5pt,after skip=5pt]
	\begin{Verbatim}[fontfamily=lmtt,fontsize=\footnotesize]
SELECT DISTINCT(record_id)
FROM TableEvents AS T, FunctionInvocations AS F
  ON T.func_id = F.func_id
WHERE T.event_type IN ('insert', 'update')
  AND F.function_name in SUCCESSOR_FUNC_NAMES
  AND F.workflow_id in WORKFLOW_IDS;
	\end{Verbatim}
\end{tcolorbox}
We execute this query on our Shop dataset to find all orders made by users who earlier browsed a potentially problematic item; the average query latency is 4.8s across five runs.

\subsection{Cost Analysis}
\label{sec:cost_analysis}

Finally, we evaluate the cost of deploying \sn to the cloud.
In Table~\ref{tab:cost_analysis}, we estimate the monthly total cost of serving the Shop workload
(Hotel and Retwis trend similarly) on GCP for \sn, OW, and a commercial FaaS platform (Google Cloud Functions) using a serverless database (Firestore).
We evaluate four different load patterns: low, medium, and high patterns with 10 QPS, 1K QPS, and 100K QPS,
plus a mixed pattern of 50\% low load, 49\% medium load, and 1\% high load.
For all systems except GCF+Firestore (Firestore is pay-per-request and we exclude its storage cost),
we provision the database cluster based on peak load and conservatively assume no DBMS scaling.
We assume OW can scale its workers and controllers to minimize cost at a given load
and \sn can scale its frontend servers.

We find that \sn minimizes cost at scale compared to other systems.
At low load, GCF+Firestore is the cheapest because it can scale to near-zero.
However, at medium and high load OW is 4.8--25.2$\times$ costlier and GCF+Firestore is 2.9--44$\times$ costlier because their high overhead means they require more resources to support the same load.
Even for mixed load, OW is 2$\times$ costlier and GCF+Firestore is 1.2$\times$ costlier
because this overhead outweighs any benefit derived from separating function execution and data management.
Therefore, \sn not only is faster and provides more features,
but is also more cost-efficient than comparable systems.

\begin{table}[t]
\resizebox{\linewidth}{!}{%
\centering
\begin{tabular}{@{}c|rrr|r@{}}
\toprule
    & {\color[HTML]{333333}} &  &  &  \\
    \multirow{-2}{*}{\textbf{System}} & \multirow{-2}{*}{\begin{tabular}[c]{@{}c@{}}\textbf{Low Load}\\\textbf{10 QPS}\end{tabular}} & \multirow{-2}{*}{\begin{tabular}[c]{@{}c@{}}\textbf{Mid Load}\\ \textbf{1K QPS}\end{tabular}} & \multirow{-2}{*}{\begin{tabular}[c]{@{}c@{}}\textbf{High Load}\\\textbf{100K QPS}\end{tabular}} & \multirow{-2}{*}{\begin{tabular}[c]{@{}c@{}}\textbf{Mixed}\\\textbf{Load}\end{tabular}} \\\midrule
    OW + VoltDB &\$1,221	&\$4,422	&\$153,956	&\$6,732 \\
    GCF + Firestore &\textbf{\$22}	&\$2,679	&\$268,380	&\$4,008 \\
    \sn + VoltDB &\$917	&\textbf{\$917}	&\textbf{\$6,099}	&\textbf{\$3,383} \\\bottomrule
\end{tabular}
}
\caption[]{Estimated monthly cost for OW, GCF with Firestore, and \sn serving the Shop workload on GCP\footnotemark, varying loads.}
\label{tab:cost_analysis}
\end{table}
\footnotetext{Prices were retrieved from the Google Cloud Pricing Calculator on 2022-12-28, using the us-west1 region as it was the cheapest: \url{https://cloudpricingcalculator.appspot.com/}.}
\section{Related Work}
\label{sec:related_work}

\noindent\textbf{Data-Centric FaaS Platforms.}
Many recent research systems seek to improve FaaS performance and functionality for data-centric applications.
We have already discussed several FaaS systems which provide transactional guarantees in \S\ref{sec:semantics}.
Similar systems include AFT~\cite{aft}, which interposes between a FaaS platform and a remote data store to enforce read atomicity, and Netherite~\cite{netherite}, which uses a reliable message queue abstraction to provide exactly-once semantics for serverless workflows,
though unlike \sn both systems physically separate functions from data.
Also related are FaaSM~\cite{faasm}, which allows functions to share memory regions, Shredder~\cite{shredder}, which provides low-latency storage functions 
by physically co-locating with a key-value store, and LambdaObjects~\cite{mast2022lambdaobjects} which co-locates FaaS storage and compute, though unlike \sn none of these supports transactions.
Orleans~\cite{orleans1,orleans2} virtual actors resemble stateful functions,
though their data model is based on local objects instead of a database
and, unlike \sn, Orleans does not provide exactly-once semantics.

Another set of relevant systems includes Pocket~\cite{pocket}, Locus~\cite{locus}, and Sonic~\cite{sonic}, which propose multi-tier cloud storage backends designed for FaaS applications.
These systems are largely designed for compute-intensive tasks on large amounts of data (e.g., batch analytics)
and trade off transactions and low latency for data storage cost.
However, this tradeoff is not suitable for our target applications  where each request accesses smaller amounts of data but demand
the low latency and strong transactional guarantees of \sn.

\vspace{1mm}
\noindent\textbf{Data Tracing for Observability.}
\sn's tracing layer is related to prior research on workflow provenance~\cite{provenancesurvey} and data provenance~\cite{howwhywhere}; for example, it uses query rewrites~\cite{arab2014generic,glavic2009perm} to capture data accesses.
Workflow provenance traces the flow of data through different modules (e.g., functions) in a larger program, but assumes each is stateless.
Data provenance traces the origin of individual data items, which models the state operations of a program,
though tracing fine-grained data provenance is out of scope for \sn.

The key challenge in tracing application interactions with data is capturing control flow information and linking it to data operations.
Most existing systems rely on manual annotation,
but some have proposed automatically capturing provenance information through kernel interposition~\cite{muniswamy2006provenance}
or dynamic analysis~\cite{murta2014noworkflow},
though this information by itself is often too low-level for users~\cite{pimentel2016yin,dey2015linking}
so it must be supplemented with information from manual annotations~\cite{angelino2010starflow,muniswamy2006provenance,mcphillips2015yesworkflow,pimentel2016yin,dey2015linking}.
Other systems have proposed automatically combining workflow and data provenance for scientific and analytics applications~\cite{amsterdamer2011putting,chirigati2012towards}, but tolerate high latencies.
\sn interposes between functions and the DBMS and leverages control flow information inherent
in its FaaS programming model to automatically capture both workflow provenance and data operations
without manual annotations.
While prior systems provide information flow control-based security for FaaS~\cite{alpernas2018secure},
and secure container-based FaaS applications by tracing system calls and network activity
(and using these to infer data movement)~\cite{alastor},
we do not know of any prior work which can produce a
similarly complete record of application interactions with data in a FaaS environment.
\section{Conclusion}
\label{sec:conclusion}

We presented \sn, a novel transactional FaaS framework for data-centric applications.
\sn physically co-locates and logically integrates function execution and data management by wrapping a distributed DBMS and its stored procedures.
It guarantees functions run as ACID transactions, provides multi-function transactions and fault-tolerant workflows, and offers advanced observability capabilities.
In addition to providing more features and stronger guarantees than existing FaaS platforms, \sn outperforms them by 2--68$\times$ on microservice workloads by reducing communication overhead.
\bibliographystyle{plain}
\bibliography{references}

\end{document}